\documentclass[preprints,article,accept,moreauthors,pdftex]{Definitions/mdpi}
\firstpage{1} 
\pubvolume{8}
\issuenum{6}
\articlenumber{303}
\pubyear{2022}
\copyrightyear{2022}
\externaleditor{Academic Editor: {
Luciano Nicastro}} 
\datereceived{22 April 2022} 
\dateaccepted{16 May 2022} 
\datepublished{26 May 2022} 
\hreflink{https://doi.org/\linebreak10.3390/universe8060303} 
\pdfoutput=1

\Title{Antarctic Survey Telescope 3-3: Overview, System Performance and Preliminary Observations at Yaoan, Yunnan}
\TitleCitation{Antarctic Survey Telescope 3-3: Overview, System Performance and Preliminary Observations at Yaoan, Yunnan}

\Author{Tianrui Sun $^{1,2}$\orcidS{}, Xiaoyan Li $^{3}$, Lei Hu $^{1}$\orcidL{}, Kelai Meng $^{1}$\orcidK{}, Zijian Han $^{3}$\orcidH{}, Maokai Hu $^{1}$\orcidM{}, Zhengyang Li $^{3}$\orcidZ{}, Haikun Wen $^{3}$, Fujia Du $^{3}$, Shihai Yang $^{3}$, Bozhong Gu $^{3}$, Xiangyan Yuan $^{3}$, Yun Li $^{3}$\orcidY{},  Huihui Wang $^{1}$, Lei Liu $^{1}$, Zhenxi Zhu $^{1}$, Xuehai Huang $^{1}$, Chengming Lei $^{1}$, Lifan Wang $^{4}$\orcidW{} and Xuefeng Wu $^{1,2,}$*\orcidX{}}

\AuthorNames{Tianrui Sun, Xiaoyan Li, Lei Hu, Kelai Meng, Zijian Han, Maokai Hu, Zhengyang Li, Haikun Wen, Fujia Du, Shihai Yang, Bozhong Gu, Xiangyan Yuan, Yun Li, Huihui Wang, Lei Liu, Zhenxi Zhu, Xuehai Huang, Chengming Lei, Lifan Wang and Xuefeng Wu}

\AuthorCitation{Sun, T.; Li, X.; Hu, L.; Meng, K.; Han, Z.; Hu, M.; Li, Z.; Wen, H.; Du, F.; Yang, S.; et al.}

\address{%
$^{1}$ \quad Purple Mountain Observatory, Chinese Academy of Sciences, Nanjing 210023, China; {trsun@pmo.ac.cn~(T.-R.S.); hulei@pmo.ac.cn (L.H.); mkl@pmo.ac.cn (K.M.); kaihukaihu123@pmo.ac.cn (M.H.);} \linebreak {wanghh@pmo.ac.cn (H.W.); lliu@pmo.ac.cn (L.L.); zxzhu@pmo.ac.cn (Z.Z.); xhhuang@pmo.ac.cn (X.H.); cmlei@pmo.ac.cn (C.L.)}\\
$^{2}$ \quad School of Astronomy and Space Sciences, University of Science and Technology of China, Hefei 230026, China\\
$^{3}$ \quad Nanjing Institute of Astronomical Optics \& Technology, National Astronomical Observatories, Chinese Academy of Sciences, Nanjing 210042, China; {xyli@niaot.ac.cn (X.L.); zjhan@niaot.ac.cn (Z.H.); zyli@niaot.ac.cn (Z.L.); hkwen@niaot.ac.cn (H.W.); fjdu@niaot.ac.cn (F.D.); shyang@niaot.ac.cn (S.Y.); bzhgu@niaot.ac.cn (B.G.); xyyuan@niaot.ac.cn (X.Y.); yli@niaot.ac.cn (Y.L.)}\\
$^{4}$ \quad George P. and Cynthia Woods Mitchell {Institute} for Fundamental Physics \& Astronomy, \linebreak Texas A\&M {University}, {Department} of Physics and Astronomy, {4242 TAMU}, College Station, TX 77843, USA; {lifan@tamu.edu}}

\corres{Correspondence: xfwu@pmo.ac.cn}
\abstract{The third Antarctic Survey Telescope array instrument {at} Dome A in  Antarctica, the AST3-3 telescope, has been in commissioning from March 2021. We deployed {AST3-3} at the Yaoan astronomical station in Yunnan Province for an automatic time-domain survey and follow-up observations with an optimised observation and protection system. The telescope system of AST3-3 is similar to that of AST3-1 and AST3-2, except that it is equipped with a {14K~$
\times$~10K} QHY411 CMOS camera. AST3-3 has a field of view of $1.65^\circ \times 1.23^\circ$ and {is currently using} the $g$ band filter. During commissioning at Yaoan, AST3-3 aims to conduct an extragalactic transient survey, coupled with prompt follow-ups of opportunity targets. {In this paper, we present the architecture of the AST3-3 automatic observation system.} We demonstrate the data processing of observations by representatives SN 2022eyw and GRB 210420B.}

\begin{document}
\section{Introduction}
In recent years, short time-scale transient sources have become a research hotspot in time-domain astronomy. The~time-varying scale of short time-scale transient sources ranges from milliseconds to days and is often associated with violent astrophysical processes, such as supernovae, gravitational wave {events, fast radio bursts}, gamma-ray bursts, high-energy neutrino events, tidal disruption events of stars.
On the one hand, the~causes and physical mechanisms of these phenomena still have {open questions}.
They are also observational probes closely related to the frontier of astronomy and physics, such as the verification of {general relativity and quantum gravity}, dark energy, dark matter, and~physics beyond the standard model of particle physics. 
On the other hand, for~transients with short time scales, especially the time-scale of milliseconds to minutes, there is still a lack of systematic observations in the optical waveband. The~study of transients requires surveys with a wide field of view (FoV), high photometric precision, and~{complete time coverage}.  {Large time-domain surveys have developed rapidly in recent years all over the world, such as the Panoramic Survey Telescope And Rapid Response System \citep{PANSTARRS19}, Zwicky Transient Factory \citep{2014htu..conf...27B,2019PASP..131a8002B}, Asteroid Terrestrial-impact Last Alert System \citep{2018PASP..130f4505T}, Gravitational-Wave Optical Transient Observer \citep{2020MNRAS.497..726G} , Skymapper \citep{2017PASA...34...30S}, Deca-Degree Optical Transient Imager \citep{2021MNRAS.507.1401B} and the upcoming Large Synoptic Survey Telescope \citep{2015salt.confE..20W}. 
In China, large field surveys are also developing rapidly such as the THU-NAOC Transient Survey \citep{2015RAA....15..215Z}, Tsinghua University-Ma Huateng Telescopes for Survey \citep{2020PASP..132l5001Z}, and~China Near-Earth Object Survey \citep{2015AcASn..56..178W}.} 

{The Antarctica continent possesses very competitive astronomical observation sites because of rare human activities and long astronomical polar nights  \citep{2021MNRAS.501.3614Y}. }
Extremely temperature ensures the water vapour content low and stable, thus reducing photometric noise caused by vapour absorption \citep{2006PASP..118..489K}.
{Atmospheric turbulence on the plateau also reduces the scintillation noise and improves photometric precision \citep{2005AntSc..17..555S,2006SPIE.6267E..1LL,2013IAUS..288...15A}.}
In addition, the~Antarctic continent has the cleanest air on Earth, with~the lowest atmospheric aerosol concentration and negligible artificial light pollution \citep{2007ChA&A..31...98S}.
Dome A is located at the highest point in Antarctica at 77.56$^\circ$ E and 80.367$^\circ$ S , with~an elevation of 4093 m ice tap, which has many advantages for astronomical observations \citep{2020RAA....20..168S,2011arXiv1101.2362Z}, such as {extreme cold temperatures and low absolute humidity} on Earth \citep{2017AJ....154....6Y,2014PASP..126..868H,2013IAUS..288..231Z}.
{The Antarctic Survey Telescope (AST3) series includes three large FoV and high photometric precision 50/68 cm modified Schmidt telescopes \citep{2012SPIE.8444E..5ML,2019RMxAC..51..135L}.
All the three telescopes focus on time-domain astronomy, including variable starts \citep{2017EPJWC.15202010W,2020ChA&A..44...41J,2020AJ....159..201L}, exoplanets \citep{2019ApJS..240...16Z,2019ApJS..240...17Z}, electromagnetic counterparts of gravitational wave events \citep{2017SciBu..62.1433H,2017PASA...34...69A}, and~other transients in sky surveys and follow-up observations. }

{AST3-1~\cite{2020MNRAS.496.2768M} and AST3-2 were installed at Dome A by the 28th and 31st Chinese National Antarctic Research Expeditions in 2012 and 2015. 
AST3-3 has been in commission since March 2021 at Yaoan astronomical station, in~Yunnan Province, China. AST3-3 has an FoV of  $1.65^\circ\times1.23^\circ$, and~is currently using the  $g$-band filter. In~future, AST3-3 will be equipped with a mid-infrared camera,} using the unique 2.4 $\upmu$m $K$-dark infrared window of Dome A to carry out the first Antarctic Infrared Time Domain Astronomy Research Program - Kunlun Infrared Sky Survey (KISS) \citep{2015IAUGA..2256923Y,2016PASA...33...47B,2012MNRAS.424...23Y,2018SPIE10702E..0OZ}.

In the past Observation Run 3 organised by LIGO/Virgo \citep{2020AAS...23511902F}, many telescopes were involved in searching for optical counterparts of gravitational wave events.
{We also searched the GW events with the telescopes of the CHanging Event Survey array at the Yaoan Observatory \citep{2018amos.confE..17Z}.}
{Some groups use widely distributed automated telescope arrays to monitor transients, such as the Burst Observer and Optical Transient Exploring System group \citep{2011AcPol..51b..16C},  T\'elescope \`a Action Rapide pour les Objets Transitoires \citep{2003Msngr.113...45B}, COATLI \citep{2016SPIE.9908E..5OW,2019ApJ...872..118B} and the Mobile Astronomical System of Telescope-Robots \citep{2004AN....325..580L}. }

In Section~\ref{sect:hard}, we introduce the weather-based protection system, as~well as the network and hardware of the data system. In~Section~\ref{sect:Obs}, we present the observation procedure and the structure of the observation list to explain the logical sequence of our observations. We also introduce the observation plan of the AST3-3 telescope for both time-domain surveys and targets of opportunities in Section~\ref{sect:Obs}. {The preliminary results, including data statistics and the observation of gamma-ray burst GRB 210420B are presented in Section~\ref{sect:analysis}, and~we conclude in Section~\ref{sect:conclusion}.  }
\section{Observation~Hardware}
\label{sect:hard}
\begin{figure}[H]
\includegraphics[width=.98\textwidth, angle=0]{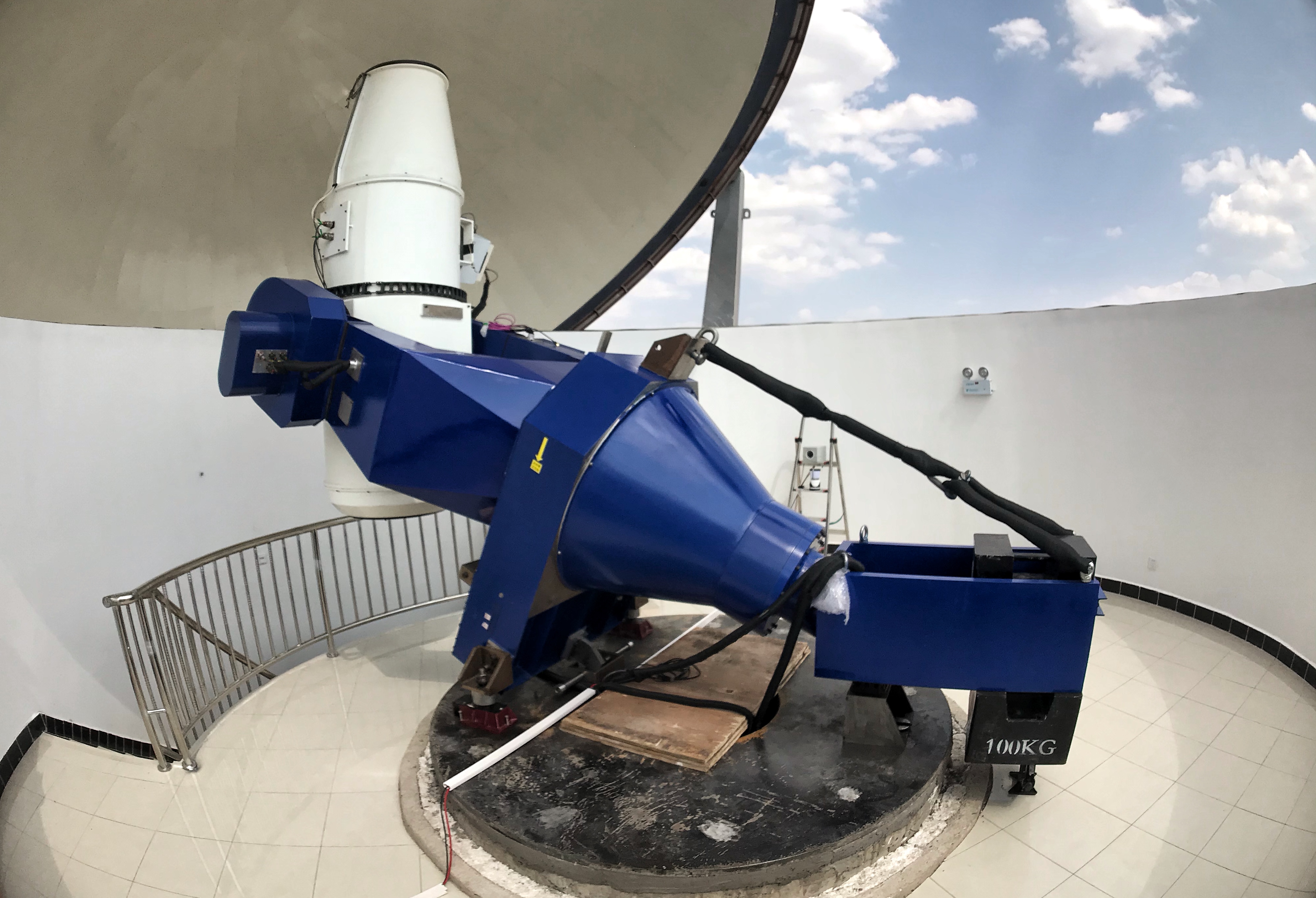}
\caption{AST3-3 deployment in~dome.}
\label{FigAST3}
\end{figure}

At the beginning of 2021, we placed the AST3-3 telescope into an eight-meter diameter dome as shown in Figure~\ref{FigAST3} at the Yaoan Astronomical Station at $101^\circ 10^\prime 47^{\prime \prime}$ E, \linebreak$25^\circ 31^\prime 43^{\prime \prime}$ N, with~an elevation of 1900 m in Yunnan Province. 
The Yaoan station has excellent logistics for supporting the long-term commissioning of telescopes. AST3-3 collaborates with YaoAn High Precision Telescope (YAHPT hereafter, \citep{2021A&A...645A..48Y}) for multiband follow-up observations of detected transient~sources. 

The pointing, focusing, and~tracking control system of AST3-3 is upgraded from AST3-1 and AST3-2 in Antarctica. 
The observation computer communicates with the telescope control computer via the local area network based on the TCP/IP protocol with private AST3-3 commands syntax. 
To avoid cable entanglement of the camera and engines,  the~reaching time-angle of $\pm180$ degrees should be avoided, which indicates that AST3-3 should not observe any object under the north polar axis. 
{With the latitude of Yaoan Station being 25.5 degrees, the~observation limit is set to 30 degrees, which can avoid bad extinction at high zenith angles and the time angle reaching 180 degrees.}

AST3-3 conducts the time-domain sky survey, follow-up observation for high energy transients and the simultaneous observation of fast radio bursts with the FAST telescope. Different modes of observation require different storage speeds and capacity performances. We built a group of computers with high-speed storage and high-performance servers, as~shown in Table~\ref{ast3Yaoanordernador}. 
The observation computer connects with the camera via optical fibres and an FPGA card for high-speed data swaps. 
It uses {a solid-state-drive with a capacity of 5 Terabytes} based RAID-50 systems to capture data as a cache to boost the I/O process, especially in the simultaneous observation of fast radio bursts {which requires} a short-exposure observation mode. The~storage server is constructed with a 220 {Terabytes} RAID array and a scalable tape library that protects the observation data, processed data and the control database. 
The pipeline server is built by a high-performance computing server with two 32-core CPUs and two Nvidia A100 (40 GiB) computational cards for data processing. The~deep-learning analysis of transient candidates and local area network web hosted in the WebPage server to avoid the crowd of the GPU time in pipeline server. All servers synchronize time to a local GPS server by the network time protocol once per~minute.
\begin{table}[H]
\caption{Computer System for AST3-3 in Yaoan~Station.}
\label{ast3Yaoanordernador}
\small
\setlength{\tabcolsep}{4.58mm}
\begin{tabular}{ccccc} 
 \toprule
\textbf{SeverID} & \textbf{Job} & \textbf{Type} & \textbf{CPU and Accelerator} & \textbf{Memory} \\ 
 \midrule
1 & Observation & R7525 & AMD {7252~$\times$~2} + FPGA & 128 GiB \\ 
2 & Pipeline &R7525 &AMD 7522 $\times$ 2 + A100 $\times$ 2 & 512 GiB \\
3 & (NFS) Storage & R740xd2 & 4214R & 128 GiB   \\
4 & Asteroid & R740xd2 & 5118 & 256 GiB   \\
5 & Web & D30 & E5-2620& 32 GiB \\
 \bottomrule
\end{tabular}
\end{table}

\begin{figure}[H]
\includegraphics[width=\textwidth, angle=0]{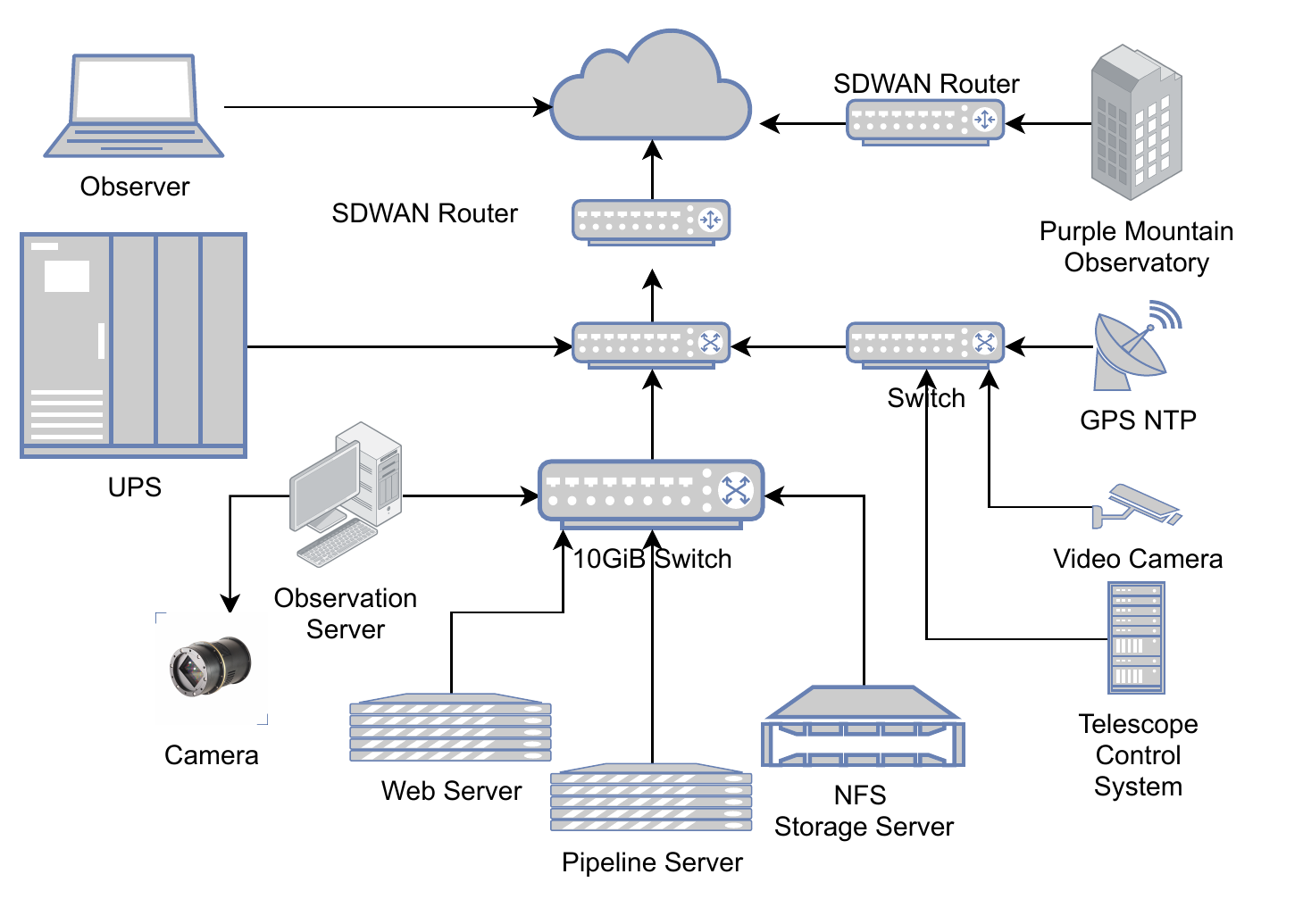}
\caption{AST3-3 private and local area Internet~structure.}
\label{FigAST3Internet}
\end{figure}

These server computers, housed in a specific room with an air-conditioner, are connected by a 10 GiB Ethernet for data swaps. The~uninterruptible power supply (UPS) controller also connects to the local area network to share the status information with all servers to save data once the power supply fails. Figure~\ref{FigAST3Internet} shows the intranet structure of the associated control schematics of AST3-3. The~software-defined wide-area network (SDWAN) is applied here for remote control and internet security for the whole system.
 
\vspace{-11pt}
\unskip

\subsection{The CMOS~Camera}
{Giving the demand of} ultrashort time-scale exposures in the observations of fast radio bursts which could increase the telescope's competitiveness, AST3-3 currently uses a complementary metal-oxide-semiconductor (CMOS) detector camera and an electronic shutter. The~AST3-3 is currently equipped with a QHY411 camera, which has a 151~megapixels Sony back-illuminated IMX411 sensor onboard and a dual thermoelectric cooler, satellite clock supports and high-speed fibre connections. This camera sensor has an exposure area of $54 \times 40$ mm$^{2}$ with a physical pixel size of $3.76 \times 3.76$ $\upmu$m$^{2}$, which contains a $14{,}304 \times 10{,}748$-pixel array. 
The FoV with this camera is $1.65^\circ\times1.23^\circ$, smaller than {those} on AST3-1 and AST3-2. The~electric shutter of this camera accepts an exposure time from 20~$\upmu$s to 3600 s, which allows both long- and short-exposure observations. The~camera provides several different exposure modes, which leads to different gain settings that would cause different system gain, readout noise, and~full-wells. 
Based on the feature datasheet of the camera, in~typical observation with readout mode \# 4, the~gain setting is usually set as 0, obtaining a system gain of 1, readout noise of 3.6 electrons and full-well near 60,000~ADUs.

The camera contains an electronic rolling shutter that introduces the feature of different start and end exposure times for each column. This difference could not occur {does not affect the physical results}
for most long-exposure observations such as supernova and afterglows of gamma-ray bursts because the difference could be ignored when comparing the shutter time to the exposure time. The~camera records the first line's exposure start time as the images' with its GPS module in short-exposure mode.

\begin{figure}[H]
 
\hspace{-8pt}\includegraphics[width=\textwidth]{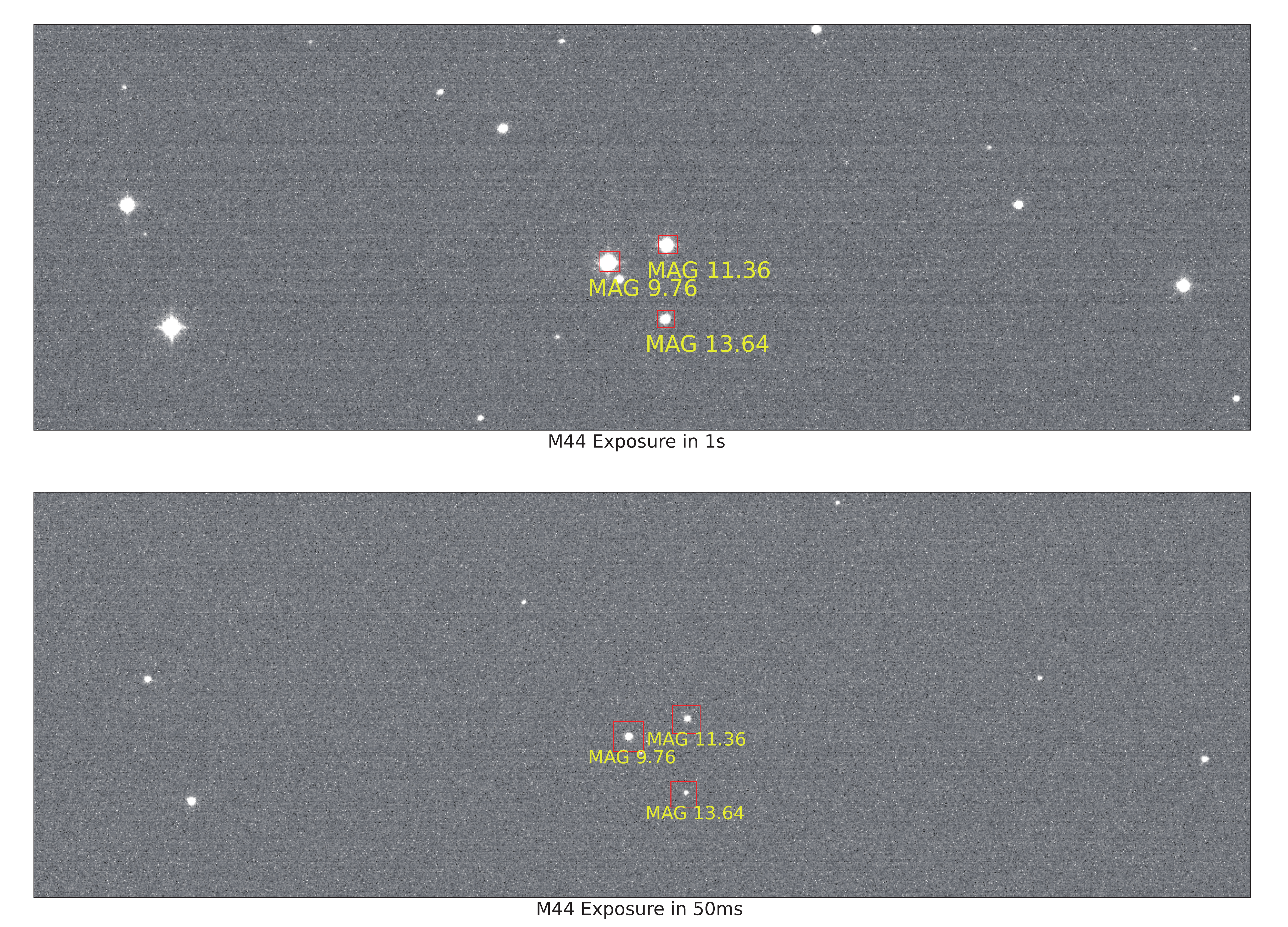}
\caption{AST3-3 Images for M44 with 1000 ms and 50 ms exposures. {The image with a one-second exposure was taken at an altitude of 65 degrees, and~the 50 ms exposure image was taken at 78~degrees. The~bright stars in the field are circled and marked with their magnitude from the SDSS DR7 {catalogue}~\citep{2009ApJS..182..543A}.} }
\label{FigCMOS}
\end{figure}

To achieve the highest frame rates for searching optical counterparts of fast radio bursts and exoplanets, we tested the maximum frame rates in the region of interest (ROI) mode with different region settings. Our pixel scale is 0.418 arcsec per pixel. The~minimally acceptable FoV is $3000 \times 1000$, with~20 arcmins by 6 arcmins. We tested the maximum frame rate and the longest exposure time at the maximum frame rate for different ROI settings and exposures with an altitude angle higher than 85 degrees on a moonless night and calculated the limiting magnitude of 5$\sigma$ in Table~\ref{Tab:fps}. 
The 50 ms mode of observation could achieve 19.6 fps and magnitude limits down to 12$\sim$13 mag in the $g$-band compared to the Pan-Starrs DR1 catalogue  (PS1, \citep{2017AAS...22923707F}).  Figure~\ref{FigCMOS} gives a comparison between a 50 ms exposure and a one-second exposure time~image. 

 
\vspace{-8pt}

The telescope is on the third floor of the dome building, and~the server computer room is on the ground floor to avoid any heat influence on the environment. To~speed up the transfer from a camera to a computer with an image data array, we applied a group of 10 Gigabytes optical fibres instead of USB wire. We use two pairs of 30-meter optical fibres for data transfer and camera control and their backup. These long fibres allow high-performance observation computers to stay away from the dome~environment.

\subsection{AST3-3 Observation~Executor}
The AST3-3 observation executes by combining an observation plan scheduler and an observation executor. The~observation executor is the main program that runs in the observation server with database interactions, camera control, and~telescope connections. More precisely, the~executor only follows the observation records in the observation table produced by the scheduler, retrieves the dome status, and~updates the camera and telescope control system status to the database. 
{The schedule table in the database is updated instantaneously by the scheduler}, which is very convenient for the requests from target-of-opportunity observations.
The schedule table contains the target name, globally unique target identifier (the task ID), right ascension, declination, priority, repeat times, and~acceptable start and end times of each~object.

\begin{figure}[H]
 
\includegraphics[width=.99\textwidth, angle=0]{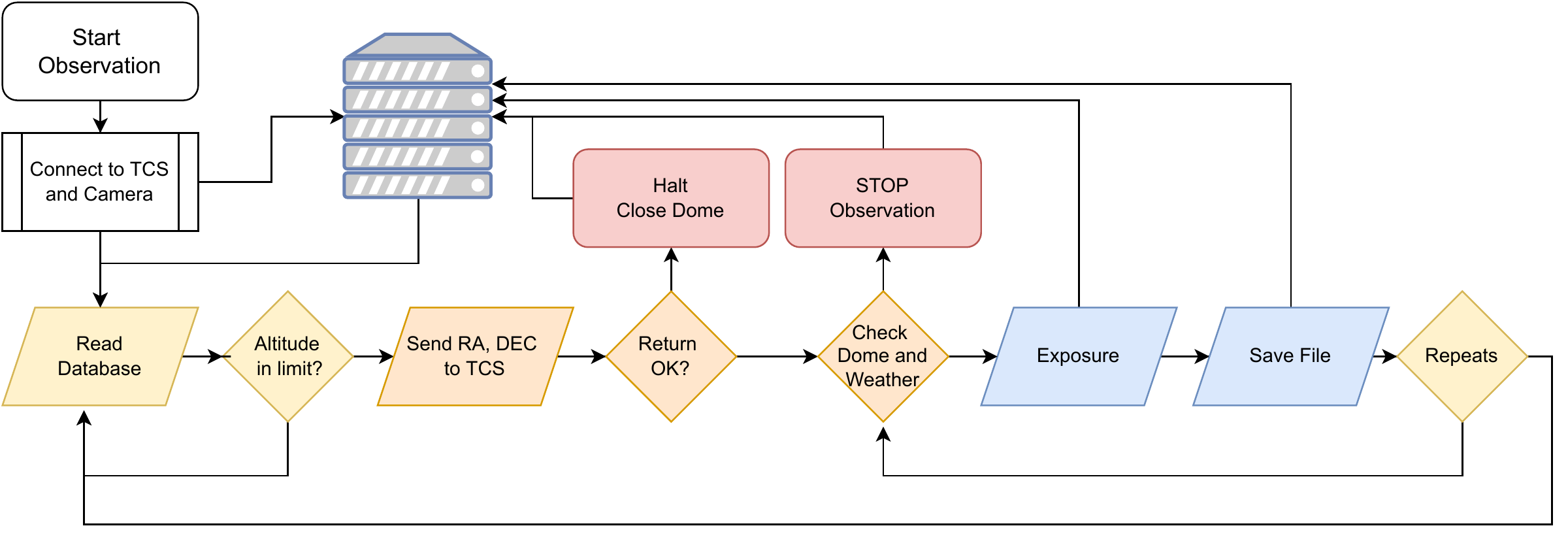}
 
\caption{AST3-3 Observation Executor workflow. The~workflow shows modules with different colours to explain their~functionality. }
\label{FigAST3logic}
\end{figure}

The executor runs by the straightforward logic shown in Figure~\ref{FigAST3logic}. For~typical observations of flat fields, the~executor starts at the Sunsets of {$-$}5 degrees altitude and stops at the Sunrises of {$-$}5 degrees. The~observation starts from the opening and connection of the camera, the~connection to the telescope control system and the focus engine's enable and placement. 
{For each observation, the~executor selects only the highest priority record with the time condition matches to be executed.}
The executor checks the observability by the altitude of the object and the distance to the moon with its coordinate by skyfield \citep{2019ascl.soft07024R} and astropy \citep{2013ascl.soft04002G}. The~object altitude should be higher than 30 degrees for each observable object by the limits from the mount and our~location.

To observe a new sky area, the~executor sends the command message with coordinates to the telescope control computer to drift the telescope to the correct position. According to the angle distances, the~span of drifting between adjacent points for stabilising is near 10$\sim$20 s. Before~each exposure, the~database supplies the dome status to the executor to ensure a fully opened dome. The~executor starts the exposure with the camera module after the telescope control system returns information as tracking.
To estimate the background immediately, the~executor calculates the background estimations for the image centre array directly after the exposure with {Python library SEP} \citep{2016JOSS....1...58B}.
The background estimation could also indicate the real-time light pollution from nearby villages and the moon in the observation logging file and the database for image~status.

The executor saves the original image data array. Some header information consists of the camera temperature before and after exposure, the~camera configuration parameters, the~humidity and pressure of the camera, the~pointing altitude and azimuth, the~telescope encoder positions, the~focus status and the background estimations of the image centre as shown in Table~\ref{Tab:head}. While saving, the~executor also uploads the header information to the image status database. If~it is a multi-image observation, the~executor will check the dome and telescope control status before each exposure. 

\subsection{Weather-Based Protection~System}

{The dry environment of Dome A means that AST3-3 works without dome protection, but~it does require a dome for protection when it runs domestically.} The dome controller should also be an automatic service for an automatic telescope to avoid humans on guard. The~YAHPT group shares their weather station data with us for the protection system of AST3-3. The~YAHPT site system has an all-sky camera, weather station and cloud station, providing weather information and critical~data.

The protection system retrieves data from the weather station once per minute to define the dome's actions. The~local weather station provides temperature, wind speed, humidity and conditional flags for clouds, wind, daylight, and~the open and close suggestions of the roof. We retrieve the weather status from the weather station and upload it to our databases by an isolated program every minute.
The dome closes when the control program meets either failure of getting weather conditions, or~bad weather on the observatory. 
This pair of guardianship programs would create a disconnection flag into the database to close the dome when the weather data {has not been} updated for more than five~minutes. 

The protection procedure works appropriately with two steps: manually or automatically open and close the dome, stop observation, and~park the telescope. When the database shows the weather anomalies or equipment failure, the~protection program sends the close dome command to the dome controller via Internet sockets. The~weather conditions for observation were restricted to wind speeds lower than ten meters per second and relative humidity lower than 70\%. Furthermore, dome closing will not affect the observation directly, and~the dome status updates every ten seconds to the database. Before~each exposure, the~observation program checks the dome status to ensure no exposures while the dome is closed. Most exposures may take 60 s, and~dome closing 
{requires a} longer time to complete, so the coincidence of exposure and closing does not {have much effect}.


\section{Basic Observation Strategy of~AST3-3}
\label{sect:Obs}
The main observational tasks of the time-domain survey are searching for supernovae, asteroids, variable stars, tidal disruption events, active galactic nuclei, and~orphan gamma bursts for AST3-3 at Yaoan station. The~target of opportunity (ToO) observation requires coordination from other facilities with high-energy transient sources such as gamma-ray bursts, gravitational wave electromagnetic counterparts, and~neutrino events. The~search area of a time-domain survey should be as large as possible to maximize the candidates. However, the~FoV of AST3-3 restricts the observable sky area each night, which requires sky area selections and a scheduler for observation. The~AST3-3 scheduler contains both the survey strategy and the follow-up observations for the target of opportunity modules. This section introduces the detailed workflow of the AST3-3~scheduler.

\subsection{Transient Survey~Strategy}
The time-domain survey compares images and catalogues among different observations, which could use fixed sky areas to facilitate the transient searching procedure. By~limiting the altitude, the~declination range is restricted to {$-$}30 degrees to 85 degrees, which allows us to use the $g$ band magnitude data from Pan-Starrs Data Release 1 as the reference for magnitude~calibrations. 

The sky area grid is slightly smaller than the AST3-3 FoV to overlap the observations on the edge. The~sky area grid also matches the Yale Bright star catalogue \citep{1987BAAS...19..733W} to avoid bright star influences on the image. Some of the lower galactic latitude sky areas may be influenced by galactic extinctions, and~we also calculate the mean galactic extinction~\cite{1998ApJ...500..525S} for each sky area. Before~the time-domain survey, the~main observation task for AST3-3 is preparing template images to optimize the image subtraction process in survey and follow-up~observations.

As the FoV is not large enough to search the entire observable sky areas,  we selected some specific sky areas as the main observation targets to reduce the cadence of the survey and enhance the light curve of observed objects. We choose sky areas with deep-field observations as shown in Table~\ref{tab:deepsurvey}, 
including sky areas from the Cosmic Evolution Survey~\citep{2007ApJS..172....1S}, 
the Great Observatories Origins Deep Survey (GOODS, \citep{2004ApJ...600L..93G}), the~Galaxy And Mass Assembly survey (GAMA, \citep{2015MNRAS.452.2087L}), 
the European Large Area ISO Survey (ELAIS, \citep{2000MNRAS.316..749O}),
and the Hyper Suprime-Cam Subaru Strategic Program \citep{2022PASJ..tmp...12A}.

We build an automatic scheduler to optimize the survey cadence and the air mass for each image each night before the observation. The~observation plan for each night is only considered in  {optimal weather conditions}  and a fixed moon phase. The~scheduler calculates the twilight flat field plan at first each night with the condition that the flat field should be half-maximum of the full-well value when the solar altitude is between $-$5 and $-$10 degrees. It selects the observable sky list by calculating the critical information on the beginning and stop times, including the local sidereal time, planet and moon positions, and~solar position. 

{Within the typical exposure time of 60 s, AST 3-3 could observe about 400$\sim$480 images in different seasons. It covers sky areas of about 800$\sim$1000 square degrees for single-visit observations.} The scheduler schedules the special sky areas for two to three observations per night. We observe special sky areas about one hour before and after transit at the meridian to achieve lower extinction. The~selected sky areas are divided into groups of ten to twenty nearby sky areas for a quick group return visit. It calculates the observation time for each target sky area in the organised sequence. It also estimates the pointing from the last target as the duration for each target with the postman sequence method~\cite{sokmen2019}. The~observation sequence is submitted to the database for observation and requested by the executor during the observation time.

\begin{table}[H]
\small
\caption{Special Sky~Areas.}
\label{tab:deepsurvey}
\setlength{\tabcolsep}{8.5mm}
\begin{tabular}{lccc} 
\toprule
\textbf{Name} &\textbf{R.A.} & \textbf{Dec.}& \textbf{Pointings}\\
\midrule
GAMA\_G02 & 30.2$\sim$38.8&	$-$10.25$\sim$$-$3.72 &	33 \\
GAMA\_G09 & 129.0$\sim$141.0 &$-$2$\sim$+3&	13 \\
GAMA\_G12 & 174.0$\sim$186.0 &$-$3$\sim$+2 &	26 \\
GAMA\_G15 & 211.5$\sim$223.5 &$-$2$\sim$+3 &	11 \\
GAMA\_G23 & 339.0$\sim$351.0 &$-$35$\sim$$-$30&	57 \\
GOODS\_N & 189.2291 &  62.2375  &	8 \\
COSMOS   & +150.1191 & +2.20583 & 4\\
Virgo\_Cluster & 186.75 &  12.71 & 38 \\
ELAIS\_N1  & 242.5041 &  54.51 & 1\\
AEGIS & 213$\sim$217&51.75$\sim$54.00& 4 \\
HSC\_SSP & - & -  & 609 \\
\bottomrule
\end{tabular}
\end{table}

\subsection{Automatic Follow up~Strategy}
The automatic follow up of essential transients is triggered by other telescopes or satellite facilities from the Internet. The~gamma-ray bursts, gravitational wave events, and~neutrino events are triggered by the GRB coordinate network (GCN) notices~\cite{2008AN....329..340B}. 
The CHIME \citep{2017AAS...22924219K} group distributes fast radio bursts coordinates via the VOevent protocol \citep[]{2017ivoa.spec.0320S}, and~the Transient Name Sever \citep{2021AAS...23742305G} distributes coordinates via email.
We have already built up a system for receiving and analysing these notices. Moreover, the~operator checks the reports from The Astronomer's Telegram \citep{1998PASP..110..754R} or the GCN Circular~manually.

The centre of the GCN distributes their machine-readable notices via a custom TCP socket protocol to boost the follow-up observations for many telescopes. Since 2012, they have distributed GCN notices in Extensible Markup Language by the protocol suggested by the VOEvent \citep{2017ivoa.spec.0320S} recommendation and allowed anonymous receivers, which could be very flexible for new observation facilities and amateur astronomers. 

\subsubsection{Gamma-Ray~Bursts}

The transient parameters include the type, coordinates and errors, trigger time, and~attachment information in most cases. Astronomical satellites and ground-based detectors aimed at the high energy transients distribute their notices via GCN notices such as the Swift \citep{2021GCN.29844....1M}, Fermi \citep{2009ApJ...702..791M, 2009ApJ...697.1071A}, GECam \citep{2020SSPMA..50l9508L}, and~IceCube \citep{2021JInst..16C0007C} projects. Hence, our scheduler immediately processes this information to the observation list with the predetermined observation~strategy.

The strategy gives each type of GCN notice its priority due to the procedure for processing and sending messages. For~example, once the SWIFT satellite triggers a gamma-ray burst by the BAT, it sends the notice immediately, and~the XRT and UVOT follow. Thus the later notice from UVOT has a more accurate position than the previous coordinate in the notice by BAT. The~Gamma-Ray Burst Monitor \citep{2009ApJ...702..791M} and Large Area Telescope \citep{2009ApJ...697.1071A}  on Fermi have different coordinate accuracies in a series of real-time and ground-generated notices when we preset separate observation plans. We evaluate and preset the priority in the observation strategy by the coordinate precision in the notice. The~preset strategy includes the exposure time, priority and repeats of exposures, which rely on the notice type. Therefore, we have built a flexible and portable system based on applicable type notices as~targets. 

An independent receiver connects to the GCN server and retrieves all notices automatically. The~scheduler reads the packet type and coordinates from the saved notice by the receiver. If~the packet type in Table~\ref{Tab:highenergytransnotices} and the coordinates supplied in the GCN notice seem observable,  it would find the matched sky areas and observe these sky areas with the presets' exposure time, priority, and~repeats. The~GCN network distributes several notices for a typical gamma-ray burst consecutively over several seconds, and~the scheduler analyses them separately to avoid procedural conflicts. It transforms each observable GCN notice into an observation request in the database with presets. The~priority in Table~\ref{Tab:highenergytransnotices} is a solution for visible object conflicts as the observer program would use the highest priority record in the~database.

\begin{figure}[H]
\includegraphics[width=10cm]{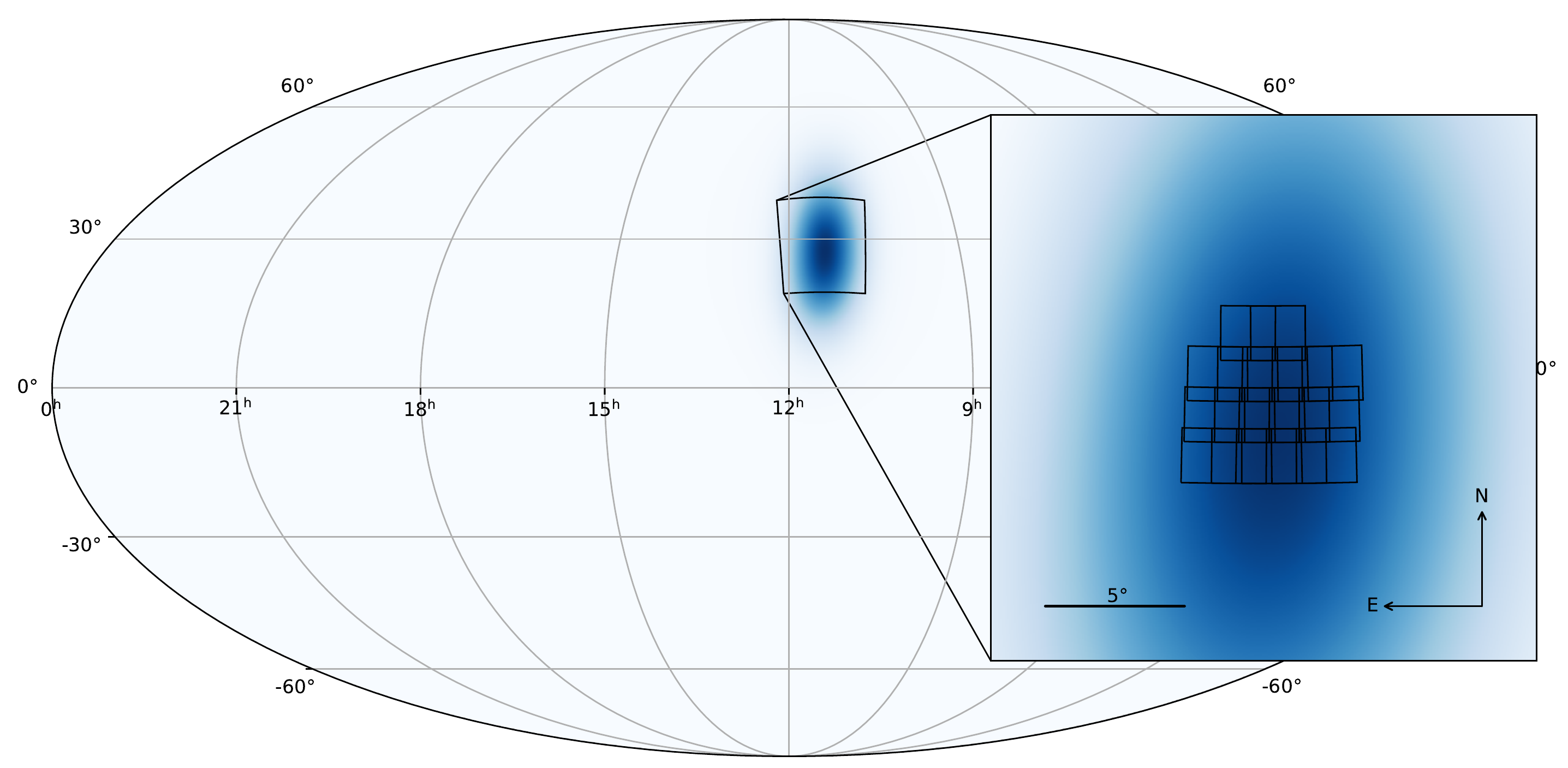}
\caption{{The target of the opportunity observation program produced} the observation grids and the observation pointing sequences for GRB 220321A. The~trigger number is 669537987, and~the probability data from GCN {Circular} \citep{2022GCN.31777....1F}.}
\label{fig:grbtootest}
\end{figure}

\begin{table}[H]
\caption{{ToO Observation Strategies.}}
\label{Tab:highenergytransnotices}
\begin{adjustwidth}{-\extralength}{0cm}
\small
\setlength{\tabcolsep}{3.03mm}
\begin{tabular}{ccccccc}
 \toprule
\textbf{ {Packet Type}} & \textbf{Name}  &\textbf{Priority}  & \textbf{RadiusErr} & \textbf{Pointings} &\textbf{Repeats} &\textbf{Exposure Time} \\
 \midrule
53 & INTEGRAL\_WAKEUP &140 & 10$^\prime$ & 1 & 30  & 60 s   \\   
54 &INTEGRAL\_REFINED  &141 & 5$^\prime$ & 1 & 30  & 30 s   \\   
55 & INTEGRAL\_OFFLINE &142 & 3$\sim$5$^\prime$ & 1 & 60  & 60 s   \\  
\midrule
61 & SWIFT\_BAT\_GRB\_POS\_ACK  &131 & 1-5$^\prime$ & 1 & 30  & 30 s    \\   
65 & SWIFT\_FOM\_OBS &132 & 10$^\prime$ & 1 & 60  & 30 s   \\   
67 & SWIFT\_XRT\_POSITION & 134 & 5$^{\prime\prime}$ & 1 & 60  & 60 s  \\   
69 & SWIFT\_XRT\_IMAGE   &135 & 5$^{\prime\prime}$ & 1 & 60  & 60 s   \\   
81 & SWIFT\_UVOT\_POS   &137 & 2$^{\prime\prime}$ & 1 & 60  & 60 s      \\   
84 & SWIFT\_BAT\_TRANS  &133 &  10$^\prime$& 1 & 60  & 60 s     \\  
\midrule
110 & FERMI\_GBM\_ALERT  &100 &4$\sim$$10^{\circ}$ & $\sim$20 & 5  & 30 s    \\   
111 & FERMI\_GBM\_FLT\_POS &101 & 4$\sim$$10^{\circ}$ & $\sim$20 & 5  & 30 s    \\   
112 & FERMI\_GBM\_GND\_POS  &102 & 4$\sim$$10^{\circ}$ & $\sim$40 & 5  & 60 s     \\   
115 & FERMI\_GBM\_FIN\_POS  &103 & 4$\sim$$10^{\circ}$ & $\sim$40 & 5  & 60 s     \\   
120 & FERMI\_LAT\_POS\_INI  &110 & 10--30$^\prime$ & 1  & 30  & 60 s   \\   
121 & FERMI\_LAT\_POS\_UPD  &111 & 10--30$^\prime$ & 1  & 60  & 60 s     \\   
127 & FERMI\_LAT\_GND   &112 & 10$^\prime$  & 1  & 60  & 60 s   \\   
\midrule
173	& ICECUBE\_ASTROTRACK\_GOLD  &121 & 0.2--0.75$^{\circ}$ & 1$\sim$2  & 10  & 60 s    \\   
174	& ICECUBE\_ASTROTRACK\_BRONZE   &120 & 0.2--0.75$^{\circ}$ &  1$\sim$2   & 10  & 60 s    \\   
\midrule
- & GECAM\_FLIGHT\_NOTIC  &105 & 2$\sim$$20^{\circ}$ & $\sim$20  & 5  & 60 s     \\ 
- & GECAM GND-BDS   &106 & 1$\sim$$10^{\circ}$ & $\sim$40 & 5  & 60 s     \\  \bottomrule
\end{tabular}
\end{adjustwidth}
\end{table}
\vspace{-3pt}
The preset strategy in Table~\ref{Tab:highenergytransnotices} is only used for immediate observation when the telescope observes typically, and~the target altitude is higher than 30 degrees.
{The positional errors in Table~\ref{Tab:highenergytransnotices} for different types of triggers are generated from the website of the GCN system. For~some gamma-ray bursts with very large radial error, we are only interested in a part of the sky area that is close to the centre.} 
The priority values are only for the comparison for the observation executor, but~are meaningless for the value itself. In~Figure~\ref{fig:grbtootest}, we also plot an example for AST3-3 observing the gamma-ray burst from Fermi GBM trigger 669537987 with the probability data from GCN Circular \citep{2022GCN.31777....1F}.

\unskip

\subsubsection{Gravitational Waves~Events}
The LIGO-Virgo collaboration detects the signal of gravitational wave events and distributes the trigger notice via the GCN network. Their GCN notices include the GraceID, packet type, alert type, false alarm rate, the~probabilities for binary neutron star merger, neutron star - black hole merger and binary black hole merger, and~the URL link to the three-dimensional probability sky map of the detected event. They use the BAYESTAR \citep{2016PhRvD..93b4013S} method to create the probability sky map rapidly and supply in the LVC\_PRELIMINARY and LVC\_INITIAL, LALInference \citep{2015PhRvD..91d2003V} more precisely in LVC\_UPDATE GCN notices. Sky maps contain 3D posterior probability distributions represented in the hierarchical equal-area iso-latitude pixelization (HEALPix) array \citep{2005ApJ...622..759G}. There would be many follow-up observation reports in the GCN circulars between the preliminary and the update notices, so we observe the LALInference manually and the automatic trigger scheduler only for the BAYESTAR sky~maps.

The HEALPix map stores probability data produced by BAYESTAR. We calculate a 3-dimensional probability calculation with the method in \citep{2016ApJ...829L..15S} and Python example code in~\citep{2016ApJS..226...10S}. {The probability density per unit volume at distance $r$ as Equation~\eqref{eqdpdv} in \citep{2016ApJS..226...10S} is used here as one component order of priority, }
\begin{equation}
\label{eqdpdv}
    \frac{{\rm{d}}P}{{\rm{d}}V}=\rho_i\frac{N_{\rm{pix}}}{4\pi}\frac{\hat{N}_i}{\sqrt{2\pi}\hat{\sigma_{i}}}\exp{-\frac{(r-\hat{\mu})^2}{2\hat{\sigma}_i}}
\end{equation}

{In Equation~\eqref{eqdpdv}, $\rho_i$ is the 2-dimensional probability in the direction at $i$-th pixel position, $\hat{\mu}_i$ and $\hat{\sigma}_i$ are the ansatz location and scale parameters, and~$\hat{N}_i$ is ansatz normalisation coefficient. The~FITS table of the HEALPix map contains $N_{\rm{pix}}$ records.}
The probability density $\frac{{\rm{d}}P}{{\rm{d}}V}$ describes the possibility of the gravitational wave event without consideration of the physical parameters of galaxies in the sky area. We optimize our method on follow-up observations with galaxy selections, with~the assumption that the short-GRB may be related to some parameter of the host galaxy \citep{2014ARA&A..52...43B}. During~LIGO/Virgo Observation Run 2 and Run 3, there are some methods for sorting galaxies by their physical parameters~\cite{2017ApJ...848L..33A,2014ApJ...795...43F,2015ASSP...40...35F}.
During the LIGO Observation Run 3, we use the Galaxy List for the Advanced Detector Era \citep{2018MNRAS.479.2374D} as the local galaxy catalogue in the CHanging Event Survey \citep{2018amos.confE..17Z} for observation. 

AST3-3 prepares to search the optical counterparts of gravitational wave events during LIGO Observation Run 4. 
We select the GLADE+ \citep{2021arXiv211006184D} catalogue as the local galaxy reference, which provides the estimated binary neutron star merger rate in each galaxy in Gyr$^{-1}$. Hence, we use the normalized multiplication of $\frac{{\rm{d}}P}{{\rm{d}}V}$ and our estimated star formation rate as the sorting basis for each~galaxy. 

The scheduler only calculates the sky area of the sum to the top 50\% HEALPix pixels, which could avoid the analysis of all pixels and reduce the calculation time. It sums up the normalized results of the galaxies in each sky area and sorts the result as the sequence of sky areas. We also build a program for retrieving the latest GCN circulars after observation to ensure up-to-date information and unique targets of the event in the whole community for checking the information as soon as possible by the~observer.

\section{Data~Collections}
\label{sect:analysis}
AST3-3 has been well equipped and fully automated for one year and has acquired extensive image data. The~image reduction and transient detection pipeline are based on the SFFT algorithm \citep{2021arXiv210909334H}. In~this section, we present the observation results with statistics and examples. The~observation capability of AST3-3 could be described by the statistics of full-width half-maximum (FWHM), and~magnitude limits in different situations. We also present the survey observations of SN2022eyw and the follow-up observations of GRB210420B with their light curves and~images.

\subsection{Observation Statistics of~AST3-3}
\begin{figure}[H]  
\includegraphics[width=0.91\textwidth]{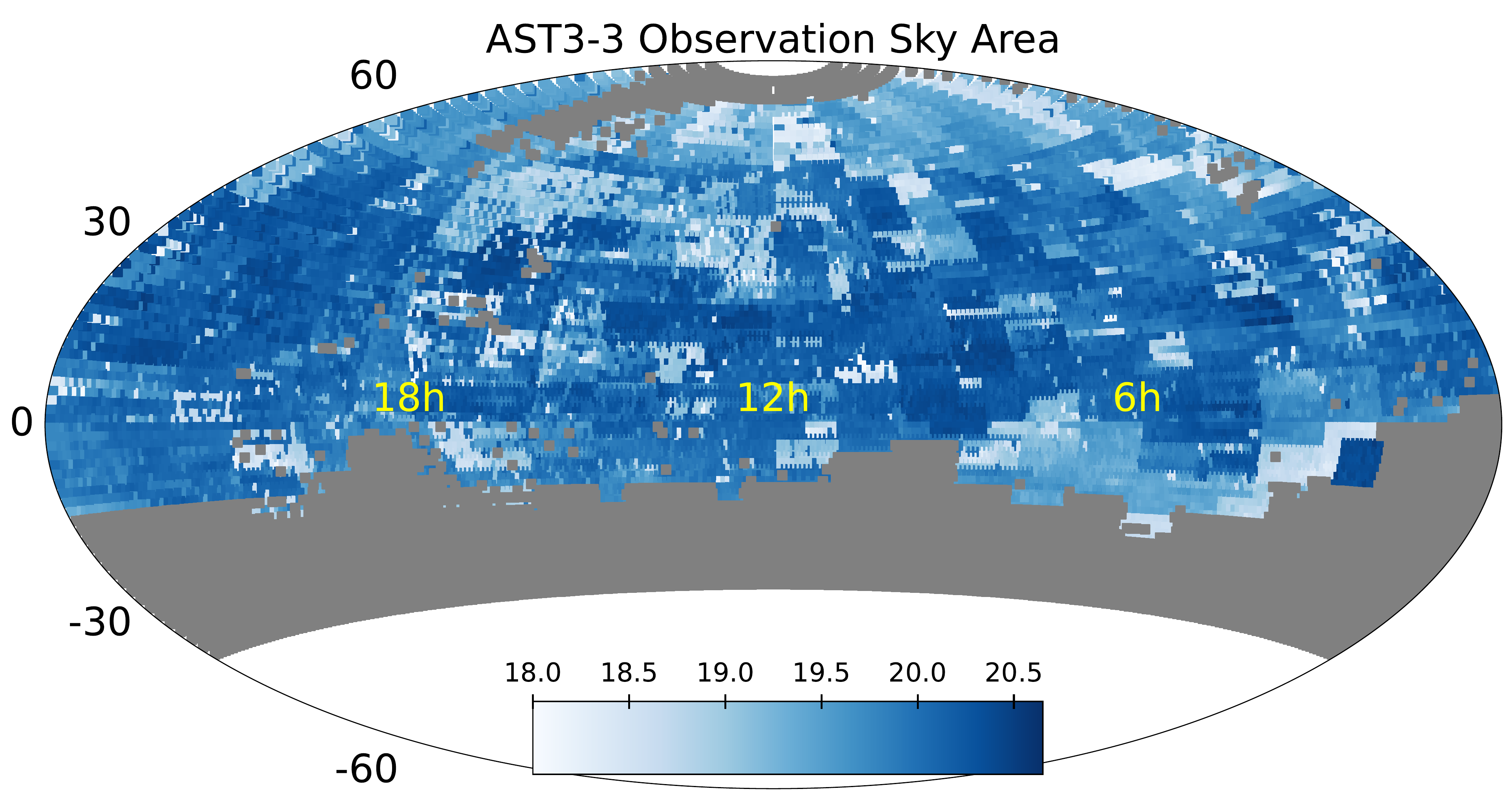}
\caption{{AST3-3 Survey Sky area grids and the magnitude} for the prepared template images till now. The~magnitude limits are coloured with depth to show that most templates are deeper than 19.0~mags.}
\label{FigAST3Survey}
\end{figure}

\begin{figure}[H]
\includegraphics[width=\textwidth, angle=0,height=200pt]{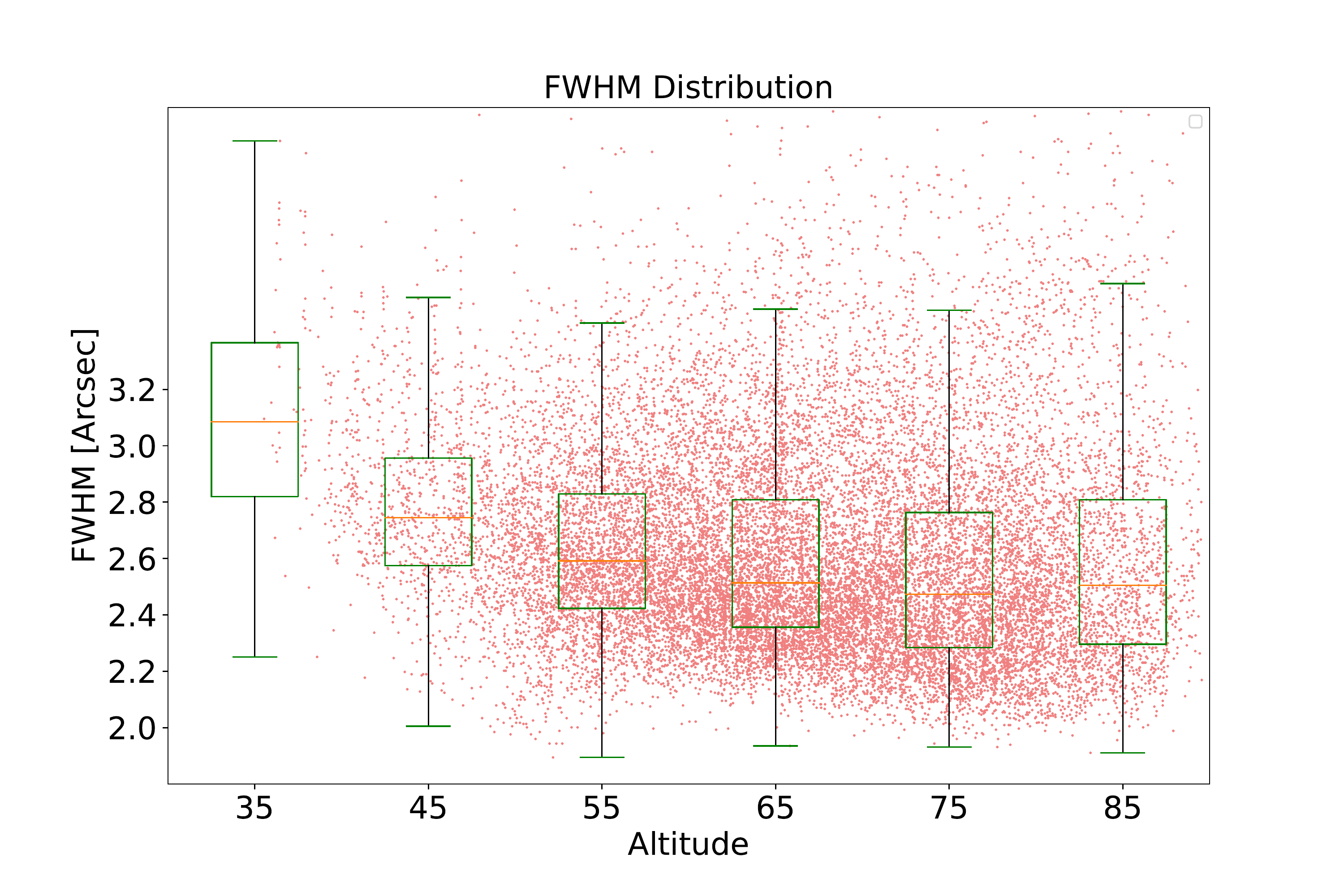}
\caption{This red points show the AST3-3 statistics for the median FWHM of images in 2021. The~box-plot shows the median and standard deviation of altitudes by ten degrees. The~FWHM becomes larger when the observation altitude is down to 45~degrees.   }
\label{FigAST3FWHM}
\end{figure}

Since March 2021, AST3-3 has captured more than 40 thousand images that have constructed a good statistical sample. AST3-3 is planned for time-domain observation, which requires the preparation of the templates. For~the northern sky, AST3-3 has taken 97\% of templates until now. Figure~\ref{FigAST3Survey} shows the sky grid of AST3-3, and~the template coverage is shown in blue points with colours for different template limit magnitudes.  
The airmass of observation affects the FWHM and the extinctions directly. Based on the stars detected on images by the source extractor \citep{1996A&AS..117..393B}, we calculated the FWHM for images. Figure~\ref{FigAST3FWHM} shows the FWHM distribution with observation altitudes. The~FWHM is centralized to 2.3 arcsec, which aligns with our~experience.
 
\vspace{-8pt}
\unskip

\begin{figure}[H]
\includegraphics[width=\textwidth, angle=0,height=200pt]{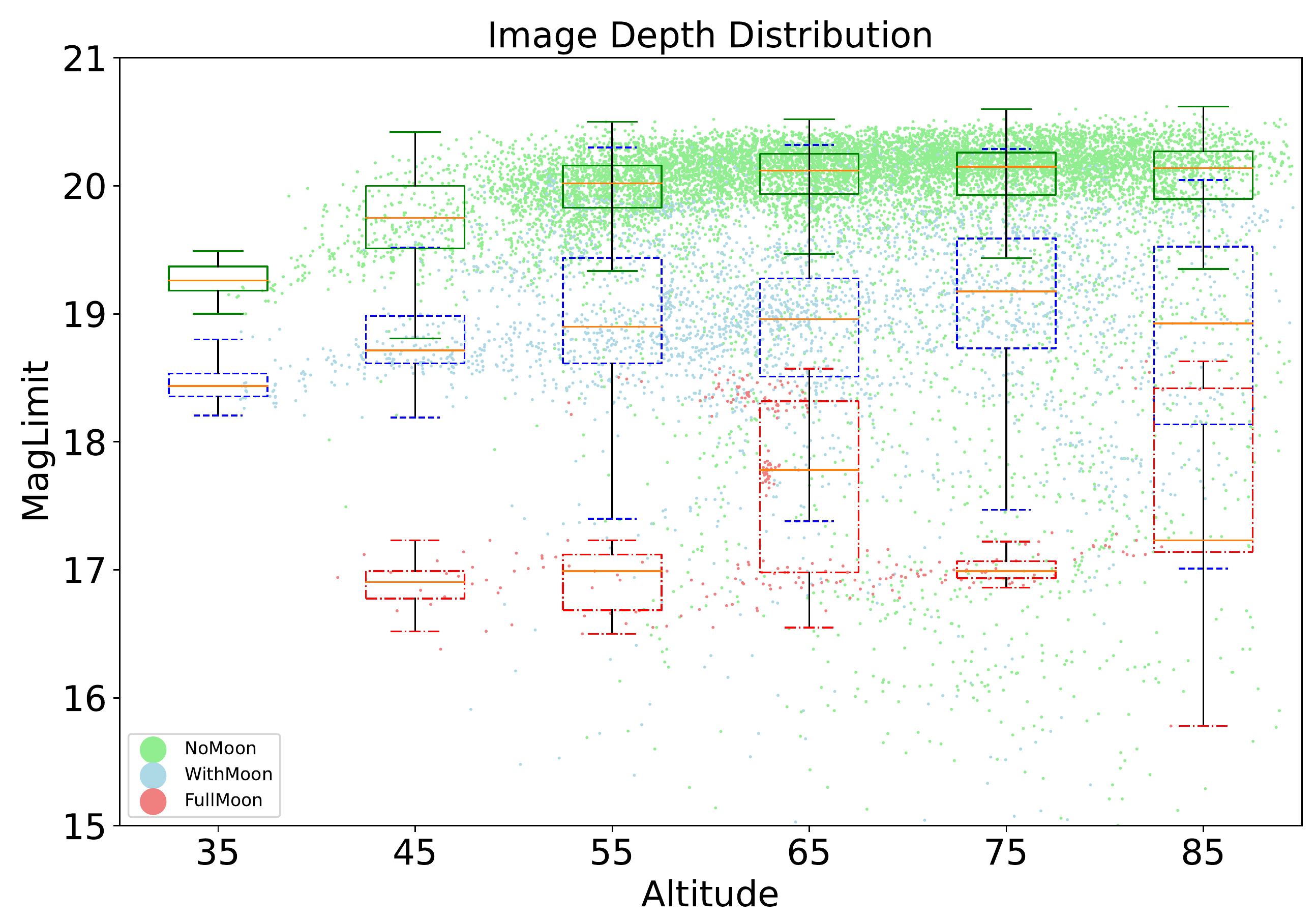}
\caption{The points shows the magnitude variation with the altitudes in different moon affections. The~box-plot shows the median and standard deviation of altitudes by ten~degrees. }
\label{FigAST3Limit}
\end{figure}

The $g$-band used by AST3-3 is strongly influenced by moonlight pollution. There is a significant difference between the limiting magnitudes with and without the moon night in Figure~\ref{FigAST3Limit}. When there is no moon in the observable sky, the~magnitude limit depends on the target's altitude. AST3-3 can obtain magnitude limits deeper than 20 mags within a 5$\sigma$ error, facilitating the search and discovery of transients. Due to the low latitudes of Yaoan Station, sometimes the Moon may have a high altitude, which reduces the magnitude limit to 16 mags or~worse. 
 
\vspace{-8pt}

\begin{figure}[H]
\includegraphics[width=.99\textwidth, angle=0,,height=200pt]{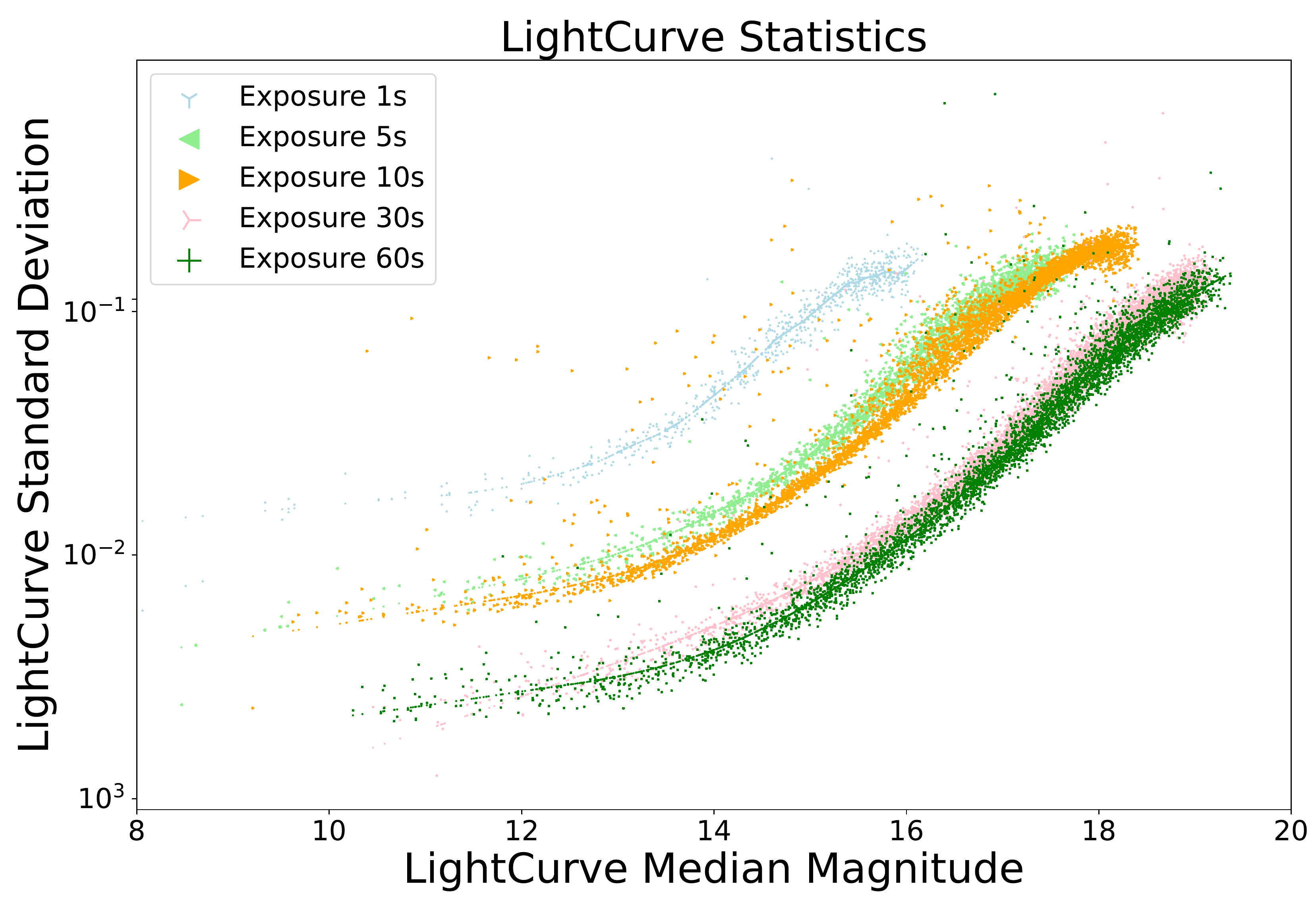}
\caption{The trends of different colour points show the magnitude statistics in light curves for stars in different exposure times from one second to 60 s. In~the centre of each exposure-time group, we use the local regression method to show the major~trends.}
\label{Fig.GRBLCs}
\end{figure}

The differential photometric accuracy is a critical feature for single-band telescopes. We analysed the differential photometry accuracy based on the standard deviation of the light curve for different exposure times. The~differential photometry of AST3-3 is performed by the vast \citep{2018A&C....22...28S} to extract the light curves with statistical parameters. Light curve statistics indicate the accuracy of the differential photometry. We plot the scattering point curve in Figure~\ref{Fig.GRBLCs} which shows the standard deviation curve with the magnitude. This curve indicates that the accuracy of AST3-3 is approximately 0.01 mag for 15.6 magnitude stars on a moonless night in Figure~\ref{Fig.GRBLCs}. 

The image coadd 
results with GRB 210420B also show the efficiency of the AST3-3 telescope. We coadd ten images to approximately 21 mags and 60 to approximately 21.5~mags with SWarp \citep{2010ascl.soft10068B}.

\subsection{SN 2022eyw Detection by the Time-Domain~Survey}

Most observation time of AST3-3 is used to construct the templates for the observable sky for transient detections in the time-domain survey and the target of opportunity observation last year. From~the beginning of 2022, the~AST3-3 survey observes special sky areas. In~this subsection, we introduce the AST3-3 observations for SN~2022eyw.

SN 2022eyw is an $i$-band magnitude of 19.66 discovered on 22 March 2022, UTC {11:04:36} by Pan-Starrs Survey and reported to the Transient Name Server \citep{2022TNSTR.762....1C}. The~coordinate of SN 2022eyw are a right ascension of 12 h 43 m 59.971 s and a declination of +62~d 19~m 48.30 s. The~AST3-3 survey {detected} this transient at a magnitude of 17.35 in the g-band from the images taken on 24 March 2022, which is part of the GOODS-N sky {areas~\citep{2004ApJ...600L..93G}. We observed the same sky area several times before the Pan-Starrs observation on 20 and 21 March 2022, but~detected nothing in the combined images as shown in Figure~\ref{Fig.SN2022eyw}.

\vspace{-8pt}
\begin{figure}[H]
\includegraphics[width=11cm]{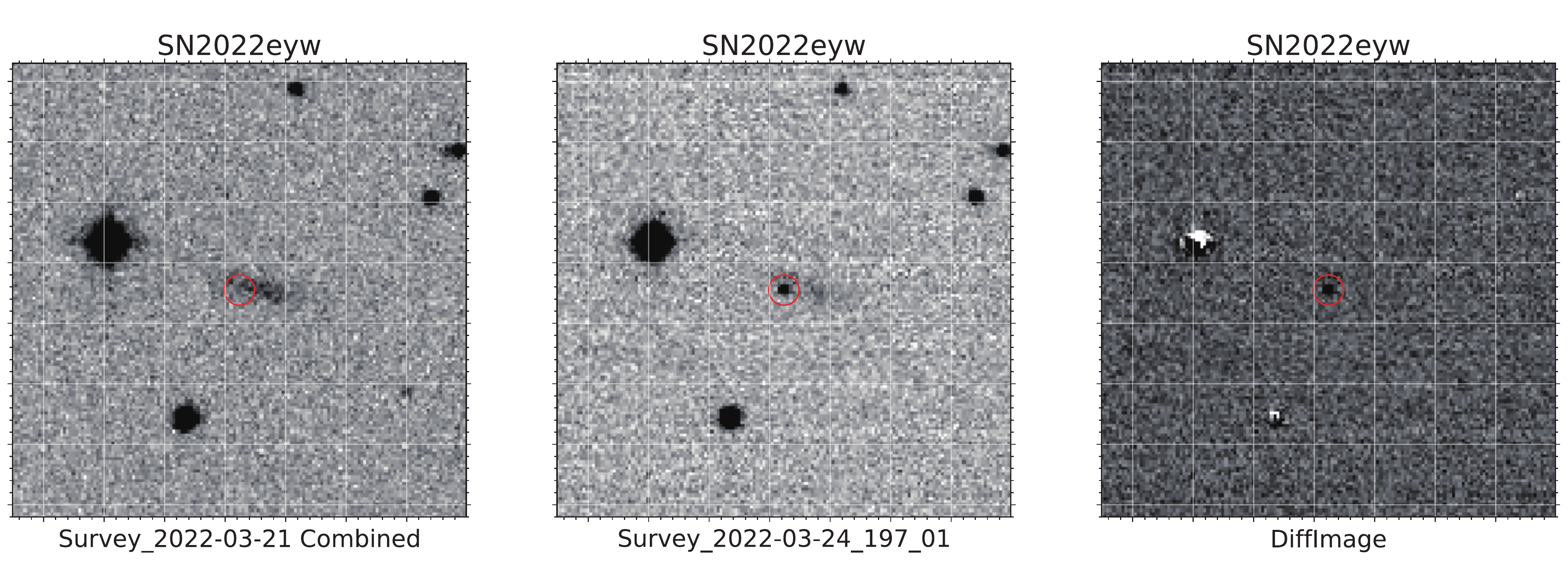}
\caption{The right image is the combination image from 21 March 2022 at the position of SN 2022eyw with a magnitude limit of 19.26. The~middle image is the first discovery image of the SN 2022eyw by AST3-3 on 24 March 2022. The~right image is the difference image produced by the transient detection pipeline with {SFFT} \citep{2021arXiv210909334H}.}
\label{Fig.SN2022eyw}
\end{figure}
Tagchi et al. \citep{2022TNSCR.783....1T} and Balcon  \citep{2022TNSCR.792....1B}
took the spectrum for SN 2022eyw and confirmed it as a rare type Iax supernova. The spectrum result was also confirmed by the Liverpool Telescope on 31 March 2022 by 
Fulton et al.  \citep{2022TNSCR.833....1F}.
From 24 March to 30 March 2022, AST3-3 acquired 28 observation images in this sky area. We plot the photometry points with data from both AST3-3 and TNS as shown in Figure~\ref{Fig.SN2022eywlc}.
\vspace{-5pt}
\begin{figure}[H]
\includegraphics[width=9cm]{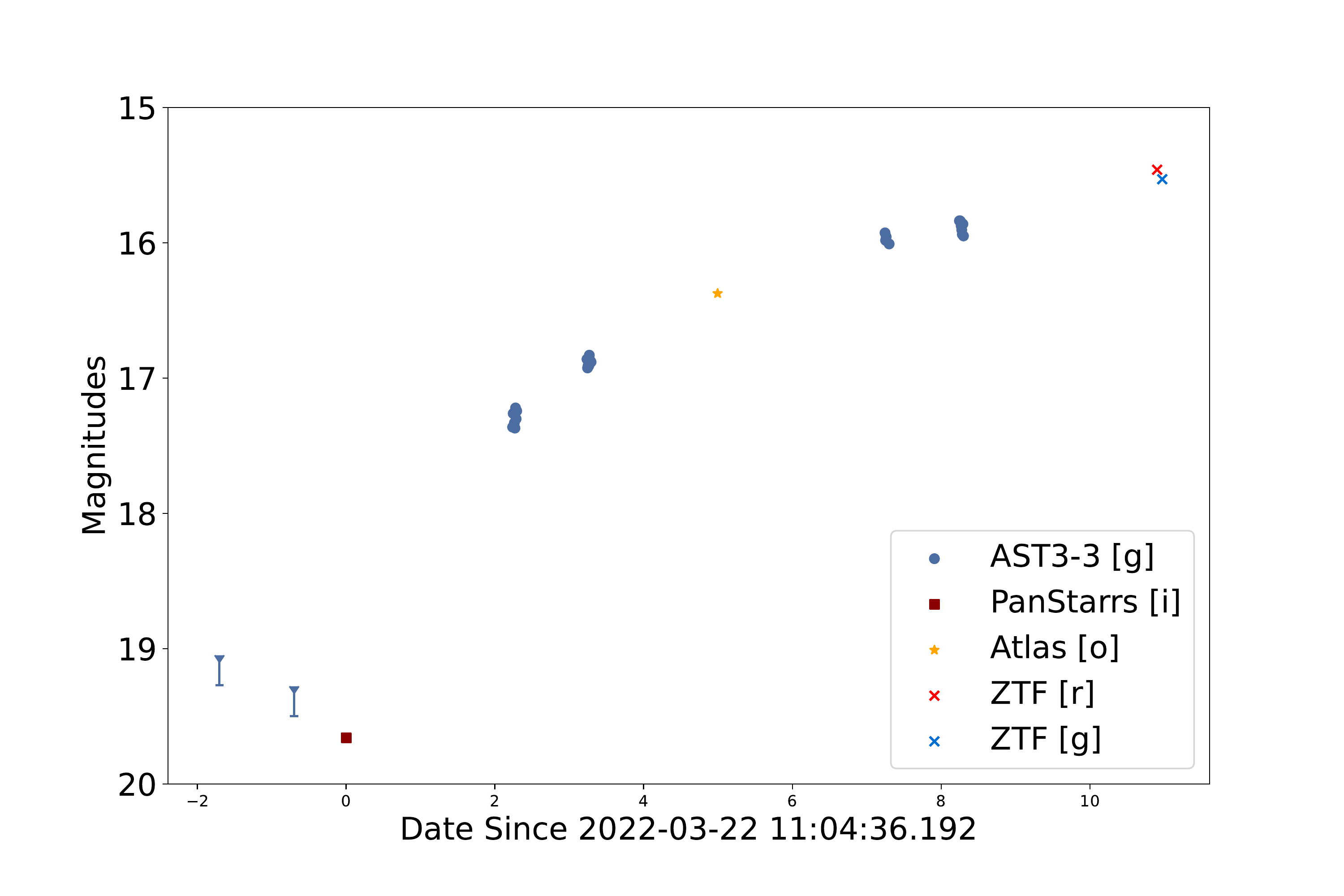}
\caption{The light curve for SN 2022eyw combines photometry data from Pan-Starrs, ATLAS and ZTF. AST3-3 observation gives the magnitude limit before the report from Pan-Starrs. The~filter o for ATLAS is orange-ATLAS. All the TNS data is taken from TNS Website. \url{https://www.wis-tns.org/object/2022eyw}  {{(accessed on 19 May 2022)}}.}
\label{Fig.SN2022eywlc}
\end{figure}
\unskip

\subsection{Follow-Up Observation of Gamma-Ray Burst~210420B}

The Swift group triggered a gamma-ray burst GRB 210420B at the position RA (J2000) of 16 h 57 m 17.39 s and Dec (J2000) of +42 d 34 m 10.2 s with an uncertainty of 3.6 arcsec~\citep{2021GCN.29844....1M}. {GRB 210420B is a long GRB with T90 (15$\sim$350 keV) of 158.8 $\pm $ 29.5 s \citep{2021GCN.29879....1S}, with~a equivalent isotropic energy of $7.7 \times 10^{51}$ erg for the redshift of 1.4. And~the Swift XRT data shows a power-law decay with an index of $\alpha$ of 2.23 and a break at T+425 s  to an index of $\alpha$ of 1.13 \citep{2021GCN.29876....1O}.} 
We began the follow-up observation with AST3-3 on 20 April 2021 at 20:07 UTC, approximately one hour and a half after the BAT trigger. We took 67 consecutive, valid images using a fixed 60 s exposure and found the object within the XRT position during this observation. {The optical afterglow of GRB210420B is shown in Figure~\ref{Fig.AST3grbImage} with coaddition to enhance the contrast of the image. There are several oscillating spikes in the AST3-3 data, which may imply the existence of voids and jumps in the ambient density of this burst~\citep{Uhm14,Geng14}. The~combined optical data shows that the optical flux decreases slowly around $10^4$ s, and~even re-brightens at $2 \times 10^4$ s. This chromatic evolution in the X-ray and the optical band indicates that a simple traditional forward shock model is not sufficient to account for all the observations \citep{Sari98}. The~late-time optical behaviour could be understood by invoking a new emission component emerging from a reverse shock produced by the additional energy injection processes \citep{Geng13,Geng16}, which will be presented in another following~work.}

The light curve of our observation is shown in Figure~\ref{Fig.AST3grbeps} with the seconds from UT 18:34:37 20 April 2020~\cite{2021GCN.29854....1L}. Our result is consistent with reports from MASTER \citep{2021GCN.29845....1L}, Hankasalmi \citep{2021GCN.29846....1O}, and~the BOOTES \citep{2021GCN.29847....1H}. Within~the candidate position \citep{2021GCN.29846....1O}, we detected a source with 18.6 mags on {20 April 2021} 20:07:30 and 19.13 mags on {20 April 2021} 21:42:43. We use the \citep{2016DPS....4812342M} method for the calculation of the magnitude zero-point. The~limiting magnitude was around 20.5 magnitude during the observation, and~we reported this information in the GCN~Circular. 

\begin{figure}[H]
\includegraphics[width=.91\textwidth]{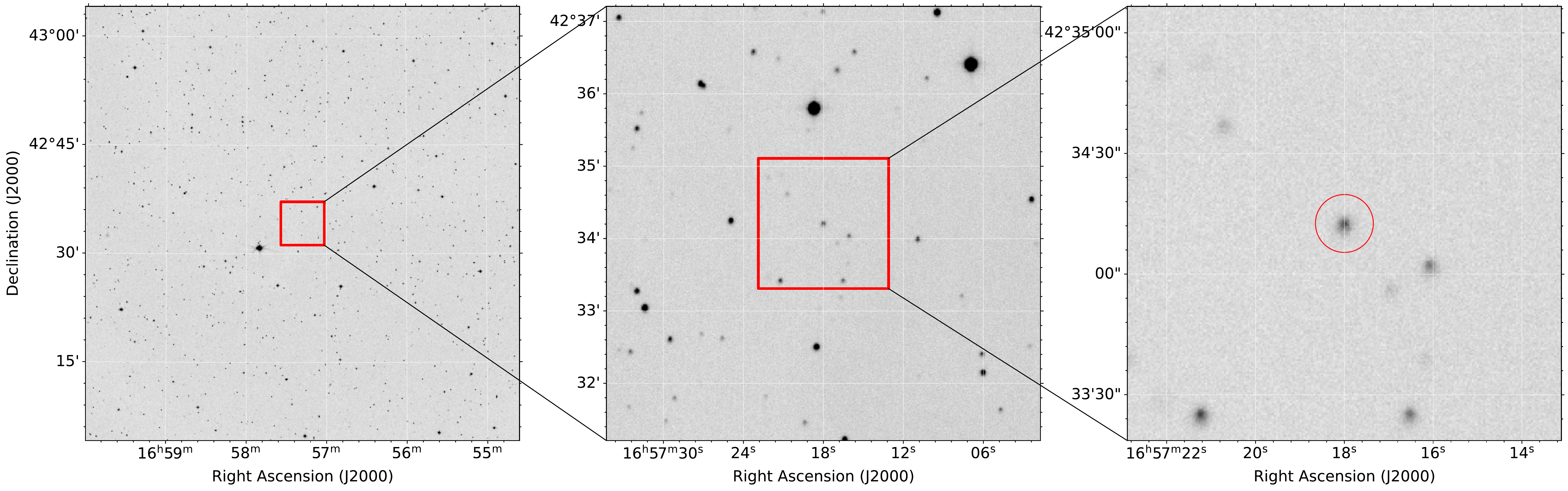}
\caption{The three panel shows the different zoom scale for the GRB 210420B images. The~\textbf{left~panel} shows a 1-degree FoV image of the GRB 210420B. The~\textbf{middle panel} shows the 0.1-degree FoV of the GRB 210420B. Moreover, the~\textbf{right panel} shows the two arcmin FoV of the GRB~210420B. }
\label{Fig.AST3grbImage}
\end{figure}

\vspace{-23pt}
\begin{figure}[H]
\hspace{-18pt}\includegraphics[width=.91\textwidth, angle=0,,height=200pt]{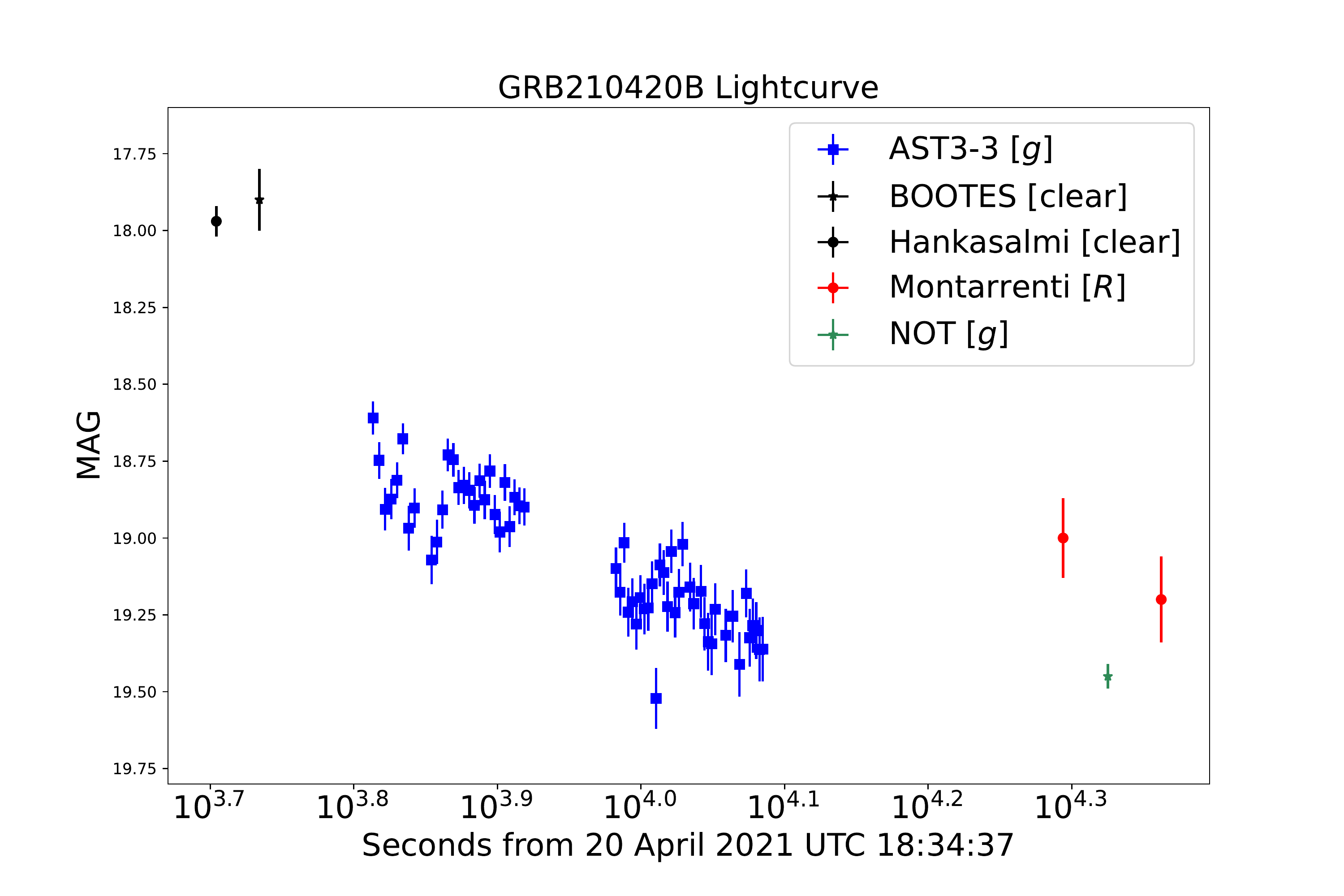}\vspace{-8pt}
\caption{{This image shows the light curve of the GRB 210420B}. The~light curve combines data from AST3-3, {BOOTES} \citep{2021GCN.29847....1H}, Hankasalmi \citep{2021GCN.29846....1O}, Montarrenti \citep{2021GCN.29851....1L} and NOT \citep{2021GCN.29852....1D}.}
\label{Fig.AST3grbeps}
%
\end{figure}
\unskip

\section{Conclusions}
\label{sect:conclusion}

This paper describes the basic information of the AST3-3 telescope, the~observation strategy, and~some of the {preliminary} results. Now, we have enabled AST3-3 to enter the data acquisition phase formally. With~the protection and network system, this telescope can fully automate the observation of every visual record in the database. {The AST3-3 telescope is a 50/68 cm aperture telescope with FoV of $1.65^\circ\times1.23^\circ$ in Yaoan, Yunnan. The~main advantage of AST3-3 in the era of time-domain astronomy is supported by its fully automated observation scheduling and data pipeline system.} Observations start and stop automatically by weather conditions, and~we only need manual checks for very special alarms due to unusual problems. We have thousands of templates for the time-domain survey. Time-domain surveys and template shooting were the main~tasks.

However, the~short-exposure observations still require more testing on short-period variable stars to test the differential photometry accuracy. Automatic follow-ups with the source coordinates from GCN Circular and Astronomers' Telegram are still another problem. The~manual recognition of the contents reduces the flexibility and extends the response time. We may need more profound~research.

\vspace{6pt}

\authorcontributions{{{Conceptualization, L.W. and X.W.}; 
methodology, T.S. and X.L.; 
software, T.S., X.L. and H.W. (Huihui Wang); 
validation, L.H. and K.M.; resources, X.L. H.W. (Haikun Wen), B.G., F.D., S.Y., L.L., Y.L., X.Y., X.H., Z.L. and C.L.; 
data curation, T.S., M.H. and L.H.; 
writing---original draft preparation, T.S. and K.M.; 
writing---review and editing, X.L., L.H. and Z.H.; 
visualization, T.S. and L.H.; supervision, X.W.; 
project administration, X.Y., Z.Z., L.W. and X.W.; 
funding acquisition, X.W. 
All authors have read and agreed to the published version of the~manuscript.}}


\funding{This work is partially supported by the National Natural Science Foundation of China (Grant Nos. 11725314, 12041306, 11903019, 12173062), the~Major Science and Technology Project of Qinghai Province (2019-ZJ-A10). The~research is also partly supported by the Operation, Maintenance and Upgrading Fund for Astronomical Telescopes and Facility Instruments, budgeted by the Ministry of Finance of China (MOF) and administrated by the Chinese Academy of Sciences (CAS). 
Tianrui Sun thanks to the China Scholarship Council (CSC) for funding his PhD scholarship (202006340174).
}


\institutionalreview{Not applicable.}

\informedconsent{Not applicable.} 


\dataavailability{Not applicable.} 

\acknowledgments{The AST3-3 team would like to express their sincere thanks to the staff of the Yaoan observation station. The~AST3-3 team would like to thank the Yaoan High Precision Telescope group for their kindly shared weather data and follow-up observation resources. The~authors thank Jinjun Geng for his wonderful analysis of the GRB 210420B data. The~authors also thank Yu-Song Cao for his kindly grammatical suggestions on our draft. The~authors are also grateful to {anonymous referees whose opinion has signiﬁcantly improved this manuscript}. This research has made use of data and services provided by the International Astronomical Union's Minor Planet Center. This work has made use of data from the European Space Agency (ESA) mission
{\it~Gaia} (\url{https://www.cosmos.esa.int/gaia}  {(accessed on 19 May 2022 )}), processed by the {\it Gaia}
Data Processing and Analysis Consortium (DPAC, \url{https://www.cosmos.esa.int/web/gaia/dpac/consortium}  {(accessed on 19 May 2022 )}). Funding for the DPAC
has been provided by national institutions, in~particular the institutions
participating in the {\it Gaia} Multilateral Agreement. This research has made use of NASA’s Astrophysics Data System Bibliographic Services. Softwares: This research made use of Astropy, \url{http://www.astropy.org}  {(accessed on 19 May 2022 )} a community-developed core Python package for Astronomy \citep{astropy:2013, astropy:2018}. The~python packages: pyephem \citep{2011ascl.soft12014R}, skyfield \citep{2019ascl.soft07024R}, matplotlib~\citep{matplotlib}, scipy~\citep{jones2001scipy}, statsmodels \citep{seabold2010statsmodels}, and~sep \citep{2016JOSS....1...58B}.}

\conflictsofinterest{The authors declare no conflict of interest.
}





\appendixtitles{no}
\appendixstart
\appendix
\section[\appendixname~\thesection]{}
\begin{table}[H]
\small
\caption{ Frame Rate For QHY411 with Optical Fibre.}\label{Tab:fps}
\setlength{\tabcolsep}{4.93mm}
 \begin{tabular}{clcl}
  \toprule
\textbf{Resolution} &  \textbf{Frame Rates at 16 Bit}   & \textbf{Max Exposure Time}  & \textbf{Magnitude Limit} \\
  \midrule
Full Frame     &   -       &  1 s & 16.59   \\  
Full Frame     &   2.0 FPS     &  500 ms & 15.73   \\  
{ {5374 Lines}}      &   3.9 FPS     &   500  ms & 15.73            \\
{3000 Lines}    &   7.2 FPS     & 135 ms  & 14.61            \\
{2000 Lines}     &   10 FPS     &   100 ms    & 13.63              \\
{1000 Lines}     &   20  FPS     &  50  ms     & 11.74          \\
\bottomrule
\end{tabular}
\end{table}
\unskip

\begin{table}[H]
\small
\caption{AST3-3 Header in Observation fits file.}\label{Tab:head}
\setlength{\tabcolsep}{6.13mm}
 \begin{tabular}{clc}
  \toprule
\textbf{Card Name} &  \textbf{Typical Value}   & \textbf{Comment}                    \\
  \midrule
CAMERA  & `QHY411MERIS' & camera ID \\
EXPOSURE&                   60 &  exposure\\
MODE    &                    4 & read out mode  \\
FILTER  &                  `{$g$}' & fixed  \\ 
DATE-OBS& `2021-06-20T17:56:51.304' & time before exposure   \\
DATE-END& `2021-06-20T17:57:53.496' & time after exposure  \\
SKY     &     923.79 & center background   \\
SKYSIG  &    12.206 & center background RMS  \\
TEMP1   &                $-$27.9 & CMOS temperature before exposure  \\
TEMP2   &                $-$27.0 & CMOS temperature after exposure  \\
PWMSET  &                255.0 & cooler PWM setting\\
COOLER  &                $-$50.0 & cooler target temperature\\
GAMMA   &                  1.0 & gamma value\\
OFFSET  &                 50.0 & offset of coms camera\\
RA      &           279.752319 & target right ascension\\
DEC     &                 10.0 & target declination\\
ALTITUDE&    74.48 & altitude \\
AZIMUTH &    178.3 & azimuth\\
SKYIDX  & `1839~+~1000' & target name\\
ENCRA   & `$-$0.518577' & encoder right ascension \\
ENCDEC  & `9.808846' & encode declination\\
FOCPOS  & `12.090' & focus position\\
HA      & `359.481423' & hour angle of target \\
\bottomrule
\end{tabular}
\end{table}




\begin{adjustwidth}{-\extralength}{0cm}

\reftitle{References}

\end{adjustwidth}
\end{document}